\documentclass[sigconf]{acmart}

\usepackage{booktabs} 
\usepackage{comment}

\setcopyright{rightsretained}

\acmConference[]{}{}{}
\acmYear{2019}
\copyrightyear{2019}

\begin{document}
\title{Search-Based Serving Architecture of Embeddings-Based Recommendations}

\author{Sonya Liberman}
\orcid{1234-5678-9012}
\affiliation{%
  \institution{Outbrain Inc}
}
\email{sliberman@outbrain.com}

\author{Shaked Bar}
\orcid{1234-5678-9012}
\affiliation{%
  \institution{Outbrain Inc}
}
\email{sbar@outbrain.com}

\author{Raphael Vannerom}
\orcid{1234-5678-9012}
\affiliation{%
  \institution{Outbrain Inc}
}
\email{rvannerom@outbrain.com}

\author{Danny Rosenstein}
\orcid{1234-5678-9012}
\affiliation{%
  \institution{Outbrain Inc}
}
\email{drosenstein@outbrain.com}

\author{Ronny Lempel}
\orcid{1234-5678-9012}
\affiliation{%
  \institution{Outbrain Inc}
}
\email{rlempel@outbrain.com}


\begin{abstract}
Over the past 10 years, many recommendation techniques have been based on embedding users and items in latent vector spaces, where the inner product of a (user,item) pair of vectors represents the predicted affinity of the user to the item. A wealth of literature has focused on the various modeling approaches that result in embeddings, and has compared their quality metrics, learning complexity, etc. However, much less attention has been devoted to the issues surrounding productization of an embeddings-based high throughput, low latency  recommender system. In particular, how the system might keep up with the changing embeddings as new models are learnt.
This paper describes a reference architecture of a high-throughput, large scale recommendation service which leverages a search engine as its runtime core. We describe how the search index and the query builder adapt to changes in the embeddings, which often happen at a different cadence than index builds. We provide solutions for both id-based and feature-based embeddings, as well as for batch indexing and incremental indexing setups. The described system is at the core of a Web content discovery service that serves tens of billions recommendations per day in response to billions of user requests.
\end{abstract}

%
%


\maketitle

\section{Introduction}
\label{sec:intro}
Over the past decade, many of the advances in the recommender systems space can be reduced to the following paradigm: embed (somehow) both items and users into some latent vector space, and serve to a user the items whose vectors have the highest inner product with the user's vector.
The above paradigm fits the various matrix factorization and latent feature models made popular by the Netflix Prize~\cite{koren2009matrix}, Factorization Machines~\cite{rendle2010factorization} and their variations like Field-Aware Factorization Machines~\cite{juan2016field}, methods such as Yahoo's OFF-Set~\cite{OffSet} and deep learning schemes whose penultimate layer - nodes, weights, sometimes both - represent or encode user and item vectors \cite{he2017neural,guo2017deepfm,wu2018starspace}.

Search engines can naturally serve as the core of recommendation services, as they can easily support many types of business logic filters, as well as similarity functions \cite{abdullah2013integrating,de2015combining,OpenSourceConnections}. In particular, the inverted-index structure in search engines is highly optimized for vector-space based information retrieval schemes, which are inherently based on efficient inner-product computations. Thus, it is quite natural to  implement an embeddings-based recommendation scoring function on top of a search engine, where item embeddings are indexed and where user embeddings are sent as the query.

The rich literature on embeddings-based recommendation models mostly focuses on the modelling approaches. 
However, the literature on how to actually integrate an embeddings-based model into a live, high-throughput
low-latency system, is rather scarce. While multiple papers mention leveraging search engines to serve embeddings-based recommendations, they do not describe how they cope with 
crucial implementation details that are at the heart of any large-scale commercial system. 

First, we note that the offline modeling component that learns and updates the embeddings is usually executed at a different cadence (and often by different teams) than indexing iterations, raising the need to somehow synchronize the query (user) embeddings to the same version as the indexed item embeddings.
The mechanics of this, in turn, depend on the 
indexing process - batch indexing, for example, requires different machinery than an incremental index.

Second, different types of embeddings require different solutions. In particular, we distinguish between identifier-based embeddings, where each item and user are directly embedded (i.e. the learning is based on user and item identifiers), and feature-based embeddings, where item and user embeddings are the result of a weighted sum of embeddings of their features. The basic matrix factorization model \cite{koren2009matrix,he2017neural} exemplifies identifier-based embeddings, whereas
\cite{rendle2010factorization,juan2016field,guo2017deepfm,wu2018starspace,OffSet} exemplify feature-based embeddings.

The contribution of this paper is that it tackles the careful details required when taking the intuitive idea of using search engines to serve embeddings-based recommendations into production at scale. We discuss a wide array of possibilities in the design space,
and propose concrete solutions to the challenges they pose. The principles described are at the core of a Web content discovery service that serves tens of billions of recommendations per day in response to billions of user requests.

The paper is organized as follows. Section \ref{sec:related} surveys related work. Section \ref{sec:reference} describes a reference architecture of a personalized recommender system based on a search core. Section \ref{sec:tax} breaks down the challenges of introducing embeddings into such architectures.
Section \ref{sec:arch} presents a concrete enhanced architecture that addresses the aforementioned challenges. We conclude in Section \ref{sec:conc}.

\section{Related Work}
\label{sec:related}
This work focuses on implementing embeddings based recommenders on top of search architectures. We first sample from the huge field of embeddings-based recommenders, and follow by reviewing recommender systems built on top of search engines.

Matrix Factorization~\cite{koren2009matrix}, made popular by the Netflix prize competition, embeds both users and the items into a latent feature space of a given dimension. Factorization Machines and their Field-Aware extension (FM)~\cite{rendle2010factorization,juan2016field} extend the basic matrix factorization model to model feature interactions. Every feature (an ID or attribute in general) has a latent vector, and the dot product of all pairs of feature latent vectors are summed together. OffSet~\cite{OffSet} uses embeddings to solve for interactions between user features and item features.

More recent techniques use deep neural nets to perform collaborative filtering (CF)~\cite{he2017neural} and to extend factorization machines~\cite{guo2017deepfm}. Another neural approach is Facebook's Starspace~\cite{wu2018starspace}, which embeds objects of different types into a common vector space. Okura et al.~\cite{okura2017embedding} use deep learning to generate embeddings for serving news recommendations. 

Search-based recommender systems are also a topic of growing popularity.
Abdullah et al.~\cite{abdullah2013integrating} show how CF can be integrated with matching-based search for product recommendations. De Pessemier et al.~\cite{de2015combining} propose a hybrid approach where the recommender system uses a search engine combined with CF to provide news recommendations. Turnbull~\cite{OpenSourceConnections} shows a concrete example of a recommendation system powered by Elasticsearch.

\section{Reference Architecture}
\label{sec:reference}
Our content recommendation service holds an inventory of millions of content items. Given a user and their current context, the service retrieves the most relevant recommendations and displays them to the user. The applicative ecosystem is comprised of  three major types of players. First, publishers that host the content recommendations widget on their web pages. Second, content providers that generate the content for our inventory of potential recommendations. The providers are in some cases the publishers themselves, and in other cases they are content marketers, or advertisers. Third, users who are exposed to, and interact with, our recommendations while reading articles on the publishers' websites.
The service aims to provide a set of engaging, personalized recommendations to each user, but while doing so it must comply with many business rules and restrictions imposed by both publishers and content providers. One simple example of a business rule is geographic targeting. This rule enables a content marketer to target an audience from a specific location, namely to ensure its content is recommended to users from that location only. Publishers may also define business rules, for example they may specify a list of content providers whose items may not be recommended within their site.

The high-scale, real-time nature of our system required an engineering solution capable of scoring the relevance of millions of potential recommendations, while applying a large number of business rules and restrictions per each request and candidate. This is done at a scale of billions of recommendation requests a day, with strict latency requirements of several tens of milliseconds per request. To fulfill these requirements, our recommendation service leverages a distributed search engine as its runtime core.

Every article in the inventory of our potential content recommendations is indexed within the search engine. We index its title, other metadata, and most importantly, its set of semantic features, extracted by natural language processing techniques, that will be used to determine its relevance to each potential user. For every user, the system maintains a profile that represents their interests. These profiles are generated by aggregating the semantic attributes of the content each user has consumed. We then translate users' implied interests, as manifested in their profile, into a search query, alongside a set of filters which enforce the business rules. While filtering is natively supported by the search engine's query language, relevance scoring is implemented by custom functions that can incorporate machine-learned prediction models.

\begin{figure}
  \includegraphics[width=\linewidth]{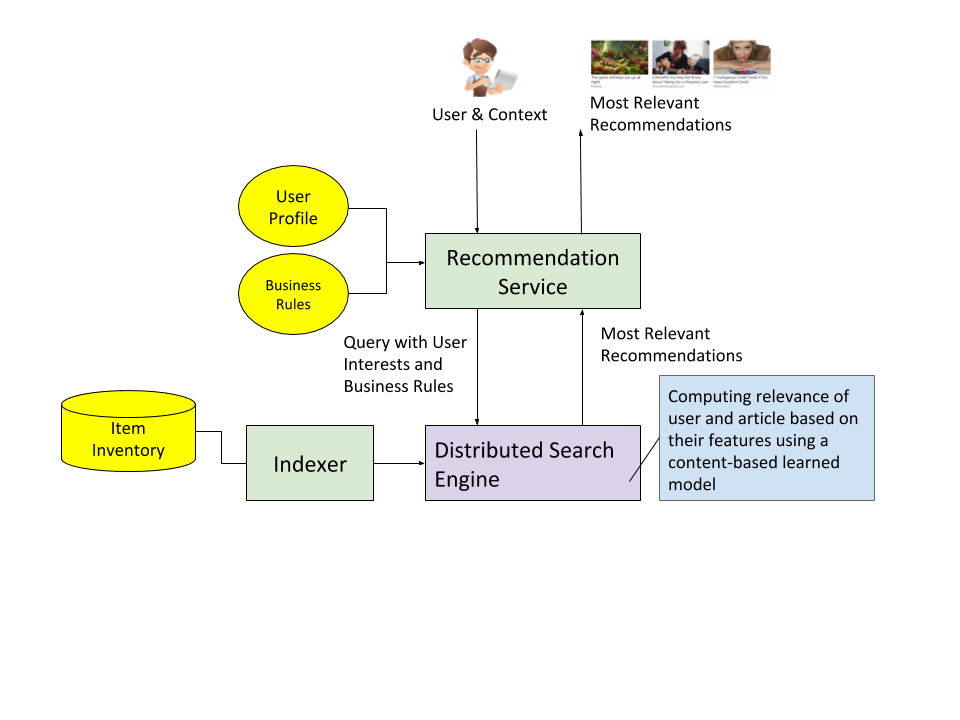}
  \vspace{-2.1cm}\caption{Reference Architecture}
  \label{fig:reference}
  \vspace{-0.1cm}
\end{figure}

Figure~\ref{fig:reference}  depicts the high-level structure of our reference architecture. The following sections describe how this architecture can be augmented to support embeddings-based recommendation models.

\section{Taxonomy of Challenges}
\label{sec:tax}
When embeddings-based models are introduced into a recommendation serving stack, a separate set of algorithmic offline flows is designed to train embeddings (potentially of several kinds) that represent users and items as dense vectors in a latent vector space.
Since new user engagement data arrives continuously, embeddings are periodically re-trained according to some frequent cadence.

We distinguish between several types of embedding models: 

{\bf Direct-Direct:} in this type of model, both user and item vectors are computed directly. Matrix Factorization techniques~\cite{koren2009matrix} are the prime example of such models, where each user and each item are projected directly into the latent space. 

{\bf Indirect-Direct:} in such a model, items are modeled directly, whereas users are modeled by the sum of the embeddings of the items they have consumed \cite{wu2018starspace}.

{\bf Indirect-Indirect:} here, both items and users are modeled indirectly, as some function of their features' embeddings. Factorization Machines and their related variants~\cite{rendle2010factorization,juan2016field} exemplify such models, as do deep collaborative representations, which combine the embeddings of user and item features using non-linear functions to derive the final user/item embedding \cite{okura2017embedding}.

Regardless of the type of the generated embeddings, when used in production, embeddings representing both the user and the items should be available in serving time.
In a search-based architecture, item embeddings  should be available within the search index along with other indexed metadata that is used for filtering and scoring candidates.
User embeddings should be passed in the search query together with other user and context data used for filtering and scoring.
Versioning and synchronization are crucial here. As aforementioned, embeddings are dynamic in their nature, and are re-computed as new engagement data is captured. Each such recomputation generates a new embedding `version'. Assessing user/item relevance when the user-side version does not match the item-side version, results in wrong or inaccurate relevance scores. Thus, the serving system must be designed such that at each point in time, the embedding version used by the query side to encode the users, matches the one indexed within the engine for all the items.

Moreover, when designing the support of embeddings-based models via a search engine, one must also consider the liveliness of the index. Here we distinguish between two modes of operation: 

{\bf Incremental Indexing:} here, the index is live and is continuously updated with new items and deletions of old 
items. Typically, updates of items are implemented as delete-and-reindex operations. 

{\bf Shadowing:} here, a read-only index serves queries, while a new index with a fresher  snapshot of the data is being built in the background. When the new index is ready, it becomes immutable and replaces the live index as the serving index. 

Section~\ref{sec:arch} presents our proposed architecture for supporting the various combinations of the aforementioned embeddings types and index liveliness modes.

\section{Embeddings-Supporting Architecture}
\label{sec:arch}
Our proposed architecture consists of three major parts: the offline embeddings training component; the indexing layer; and the serving layer. These three parts operate independently of each other, each at a different cadence while being oblivious of the others' state. 

The key new component we introduce for coordinating between the aforementioned areas is 
a stand-alone service named {\em Embeddings Orchestrator} (EO). It is responsible for exposing the embeddings, generated by the offline training component, to the indexing and serving layers, while encapsulating both the aggregation and the synchronization logic.
Section~\ref{sec:components} describes the data generated and consumed by each part of the system, followed by details on how the EO implements the aggregation and synchronization logic. Section~\ref{sec:indexing} explains how the architecture can support both shadowing (batch indexing) and incremental indexing architectures.

\subsection{Components of the System}
\label{sec:components}
The offline embeddings training component operates in cycles, and periodically generates new embeddings based on constantly incoming new training data. This layer may train one or more types of embeddings, where an {\em embedding type} is a combination of an algorithm with a specific configuration (set of hyperparameters). 
Every embedding is further associated with a {\em training time frame}, the data time frame over which it was trained. The {\em version} of each embedding vector is thus defined as a combination of the embedding type and the training time frame. Once computed, embedding vectors are stored in a distributed key-value store, one record per vector. They key holds the version of the embedding along with the identifier of the embedded entity (a specific user, item or feature), while the value holds and the vector itself. 
The offline training component and the Embeddings Orchestrator do not interact directly. 
Further below we elaborate on how the EO loads the embeddings periodically, exposing them to the indexing and serving layers.

The indexing layer is responsible for indexing the vectors of the most recent versions of the item-side embedding (perhaps of multiple embedding types) into the search index. Both the vector and its version are indexed. It obtains item embeddings from the EO, which uses the key-value store as its persistence layer.
In the Direct-Direct or Indirect-Direct embedding models, the most recent embeddings of the desired type(s) are retrieved from the EO per item-id. In the Indirect-Indirect model, the indexer will retrieve item embeddings from the EO using a list of item attributes and their weights. The EO is responsible for generating the item embedding vector by performing the weighted aggregation of the attributes' embeddings.  
It should be noted that not all items are guaranteed to have embeddings of each type and version. For example, in direct-direct models, embeddings may only be created for items that have been interacted with more than a certain number of times. 

The serving layer is responsible for fetching recommendations per user request. It first obtains the user's embedding vector, and then sends it via a search query to be scored against the indexed item embeddings. Similar to the indexing layer, the serving layer retrieves the user's embedding vector from the EO. In a direct-direct model, it retrieves the vector using user-id. In the Indirect-Direct or Indirect-Indirect models, a list of user attributes and their weights is passed to retrieve a single unified embedding. Once the embedding is returned, it is injected into the query sent to the search engine. Along with the embedding vector itself, the serving layer sends the version of the vector. The scoring mechanism ensures that the corresponding indexed item-side embeddings are used. Relevance scores for items that do not have matching embeddings are computed by a fallback scoring mechanism.

\begin{figure}
  \includegraphics[width=\linewidth]{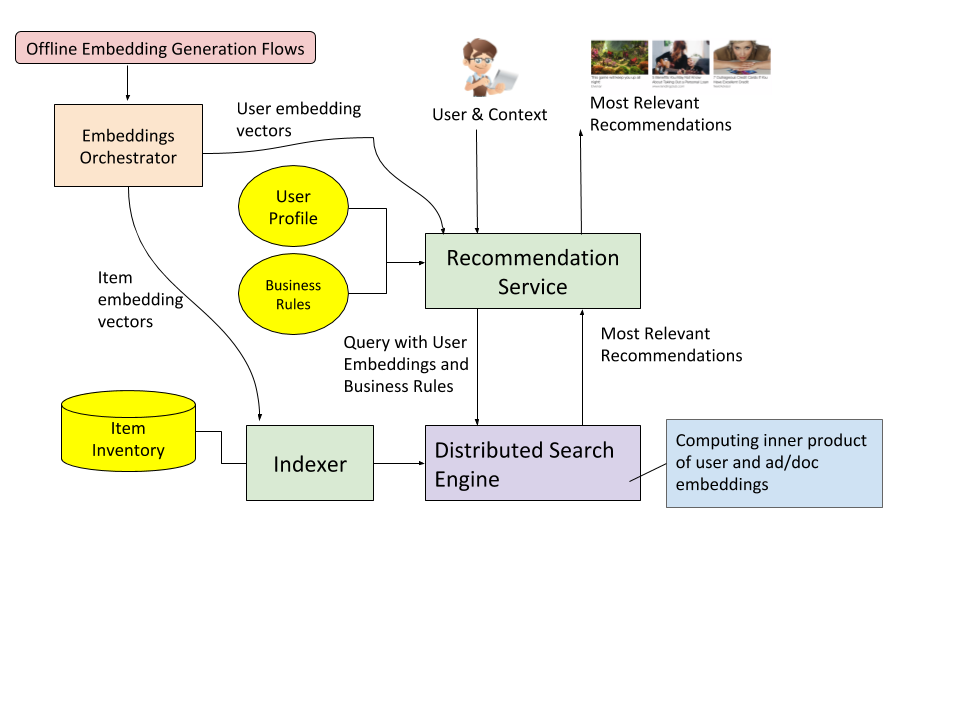}
  \vspace{-2cm}
  \caption{A Serving Architecture that Supports Embeddings}
  \label{fig:boat1}
  \vspace{-0.2cm}
\end{figure}

The Embeddings Orchestrator thus has multiple functionalities, connecting all three aforementioned layers:

{\bf Freshness of Embeddings}
The EO is responsible for periodic loading and caching of the embedding vectors created by the offline embeddings training component. For every embedding type, the EO holds a 'latest version' state, which it maintains by periodically polling the key-value store for newly generated embeddings.

{\bf Version Synchronization}
When serving with a search-based platform, it is crucial to ensure
that the embeddings versions of the indexed item vectors match that of the queried user vector. The key for such synchronization is for the serving layer to start using a new, more recent user embedding version, only after this version has been indexed for all items. This contract is enforced by the EO as follows. For each embedding type, the EO maintains a 'version in-use' state. Updates of this state are triggered exclusively by the indexing layer once all its items are indexed with the latest version. When the serving layer calls the EO to obtain an embedding for the current user, it specifies the embedding type only, without the version. The EO consults its internal 'version in-use' state, and returns the embedding vector of that 
version. Thus, the contract between the serving layer and the indexing layer is maintained, and version matching is guaranteed. Since newly generated embeddings 
only affect the 'latest-version' state of the EO but not its 'version in-use' state, new embeddings can be created in parallel to the indexing cycle without interfering with synchronization. 

{\bf Supporting Indirect Embeddings}
The EO implements the aggregation logic that combines attribute embedding vectors into a single vector that represents the user or the item. When users are embedded indirectly, the serving layer sends a list of (weighted) user attributes to the EO, which in turn retrieves the attribute embeddings and aggregates them in a weighted manner. 
Indirect item embeddings are similarly handled between the indexer and the EO.

\subsection{Incremental vs. Shadowing Indexing Architectures}
\label{sec:indexing}
In a shadowing architecture, a new index, with a fresher snapshot, is built in the background while the live index is serving queries. When the new index is ready, it becomes immutable and replaces the live index as the serving index. In each such indexing cycle, the indexer 
interacts with the EO as follows per embeddings type:
\begin{itemize}
\item It retrieves that type's 'latest-version' state from the EO.
\item For each item, it calls the EO with item id / item attributes and the latest version, to obtain an embedding vector. It then Indexes the vector and version, along with the rest of the item's data to be used for filtering and scoring.
\item Triggers the EO to set the 'in-use' version with the value of the latest version.
\end{itemize}

In an incremental indexing setting, the index is continuously updated 
in the background, while 
also serving queries, in one of two typical manners. The indexer either consumes change events from a queue in a truly continuous manner; or wakes up at short intervals to ingest a mini-batch of required changes. The description below assumes the latter model of mini-batch update cycles, but it is equally suited to the former model of consuming changes from a queue. In either case, we consider an item as "changed" (i.e. requires re-indexing) when a new embedding has been computed for it.
The mini-batch indexing cycle must ensure that all items are updated with valid embeddings that will be able to match against the user embeddings that will be sent in by the serving layer. Recall that the serving layer always sends the user's embedding vector matching the EO's 'version in-use' state. However, we want to be able to advance the 'in-use' state to equal the 'latest-version' when fresher embeddings become available. We therefore take the approach of indexing {\bf two} embedding versions per item in the incremental setting, as follows:
\begin{itemize}
\item Indexer retrieves both 'latest-version' and 'version in-use' from the EO.
\item For each item to be added or reindexed, it (1) obtains from the EO the embedding vectors of both versions; and (2) indexes both vectors, each with its own version, in addition to the other data indexed for the item used for filtering and scoring.
\item Triggers the EO to set the 'version in-use' with the value of the latest version.
\end{itemize}
This flow ensures that at each point in time, all items have a valid 'in-use' version indexed for them. 

\section{Conclusions}
\label{sec:conc}
This paper delved into the underexplored details of using embeddings in a recommendation system that uses a search engine as its runtime core. Previous work has shown that (1) embeddings are very useful in recommendation algorithms, with the inner product of an item vector and a user vector being a meaningful ranking feature; and (2) search engines are useful as the runtime core of recommender system. Furthermore, inverted indices are highly optimized for inner product operations, hence search engines can naturally support embeddings-based recommenders. However, the lack of synchronization (by design!) between index updates and the periodic recomputation of embeddings poses challenges in productizing such architectures. To overcome these, we introduced a new component called the Embeddings Orchestrator, which is responsible for synchronizing between the embeddings computation process, the indexing process, and the serving layer. 
The architecture supports direct and indirect embedding models, as well as incremental and shadowing indexing schemes. It is implemented in a large-scale online discovery service that serves tens of billions of  recommendations per day.

\bibliographystyle{ACM-Reference-Format}
\bibliography{sample-sigconf}

\end{document}